\documentclass{article}
\usepackage{spconf,amsmath,graphicx}
\usepackage{bm} 							
\usepackage{hyperref} 					
\usepackage{url}							
\usepackage{subcaption} 				
\usepackage{amssymb}
\usepackage{mathtools} 					
\usepackage{multirow} 					
\usepackage{hyperref}
\usepackage{booktabs} 					
\usepackage{makecell} 

\newcommand{\D} {\mathcal{D}}

\title{LDNET: UNIFIED LISTENER DEPENDENT MODELING IN MOS PREDICTION FOR SYNTHETIC SPEECH}
\name{Wen-Chin Huang$^1$, Erica Cooper$^2$, Junichi Yamagishi$^2$, Tomoki Toda$^1$}
\address{$^1$Nagoya University, Japan\\$^2$National Institute of Informatics, Japan}

\begin{document}
\ninept
\maketitle

\begin{abstract}
An effective approach to automatically predict the subjective rating for synthetic speech is to train on a listening test dataset with human-annotated scores. Although each speech sample in the dataset is rated by several listeners, most previous works only used the mean score as the training target. In this work, we present LDNet, a unified framework for mean opinion score (MOS) prediction that predicts the listener-wise perceived quality given the input speech and the listener identity. We reflect recent advances in LD modeling, including design choices of the model architecture, and propose two inference methods that provide more stable results and efficient computation.
We conduct systematic experiments on the voice conversion challenge (VCC) 2018 benchmark and a newly collected large-scale MOS dataset, providing an in-depth analysis of the proposed framework.
Results show that the mean listener inference method is a better way to utilize the mean scores, whose effectiveness is more obvious when having more ratings per sample.

\end{abstract}
\begin{keywords}
MOS prediction, speech quality assessment
\end{keywords}
\section{Introduction}
\label{sec:intro}
Automatic synthetic speech quality assessment is attractive owing to its ability to replace the reliable but costly subjective evaluation process. Conventional objective measures designed for telephone speech not only require a clean reference speech but also fail to align with human ratings for more varieties of speech synthesis beyond speech codec. Therefore, non-intrusive statistical quality prediction models have received increasing attention in recent years. They are typically trained on a large-scale crowd-sourced listening test like the Blizzard challenge (BC) \cite{bs2020} or the voice conversion challenge (VCC) \cite{vcc2016, vcc2018, vcc2020}), which contains speech samples and their corresponding subjective scores. Early works tried to condition simple statistical models like linear regression with carefully designed hand-crafted features \cite{linear-reg-BC2012}, while recent works use deep neural networks (DNNs) to extract rich feature representations from raw inputs like magnitude spectrum, resulting in high correlations in both utterance-level and system-level \cite{mosnet}.

Most previous works trained the mean opinion score (MOS) prediction model on the utterance-level scores. Specifically, given an input speech sample, the model was trained to predict the arithmetic mean of the several ratings from different listeners\footnote{In some literature, the term ``listener'' is also referred to as ``judge''.}. As pointed out in \cite{mbnet}, one serious problem raised by using such a training strategy is the data scarcity. Although DNN-based models require a large amount of data, due to budget constraint, only a limited number of samples from each system are rated. As a result, the number of per-utterance scores can be too small for DNN models. Researchers have tried to address this problem by pretraining on an artificially distorted dataset \cite{siamese-mos} or utilizing self-supervised speech representations (S3Rs) trained on large-scale unlabeled datasets \cite{s3r-mos}.


A more straight-forward approach is to leverage all ratings w.r.t. each sample in the dataset. This is called listener-dependent (LD) modeling, and has been studied in the context of speech emotion recognition \cite{ld-ser}. In addition to enlarging the data size, another advantage of LD modeling is more accurate modeling of the prediction by taking into account the preference of individual listeners. In the field of MOS prediction, a recent study proposed the so-called mean-bias network (MBNet) \cite{mbnet, s3r-mos}, which consists of a mean subnet that predicts the utterance-level score of each utterance and a bias subnet that predicts the bias (defined as the difference between the mean score and listener score). During inference, given an input speech, the bias net is discarded and only the mean net is used to make the prediction.

In this work, we propose a unified framework, LDNet, that summarizes recent advances in LD modeling. LDNet directly learns to predict the LD score given the input speech and the listener ID.
We also proposed two new inference methods. The \textit{all listeners} inference averages simulated decisions from all listeners in the training set, and is shown to be more stable than the \textit{mean net} inference. The \textit{mean listener} inference mode relies on a learned virtual mean listener for fast prediction.
We also suggest a more light-weight yet efficient model architecture design.
We conducted systematic experiments on two datasets, including the VCC2018 benchmark and a newly collected dataset \cite{BVCC}.
Experimental results demonstrate the effectiveness of our system, meanwhile shedding light on a deeper understanding of LD modeling in MOS prediction.

\begin{figure*}[t]
	\centering
	\includegraphics[width=\textwidth]{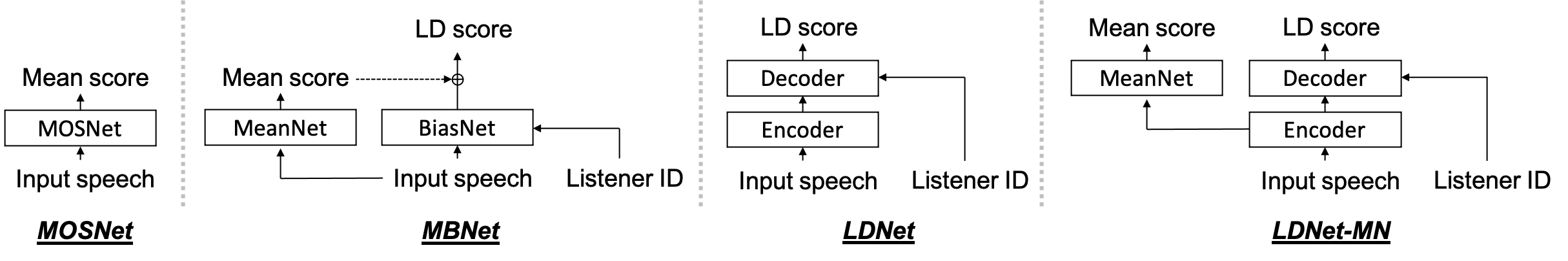}
	\centering
	\captionof{figure}{Illustration of the models described in this work. From lest to right: MOSNet, MBNet, LDNet, LDNet with MeanNet multitask learning (LDNet-MN).}
	\label{fig:all_models}
\end{figure*}

\section{Preliminary}

\subsection{Formulation}

We first introduce the problem formulation of MOS prediction modeling as described in \cite{mbnet}. Assume we have access to a MOS dataset $\D$ containing $N$ speech samples. Each sample $x_i$ has $m$ \textit{LD scores} $\{s_i^1,\cdots,s_i^m\}$ rated by a set of random listeners $\{l_i^1,\cdots,l_i^m\}$. We can further denote the \textit{mean score} as $\bar{s_i}$. Note that the same listener may have rated several different samples. In total, there are $M$ listeners in $\D$, and usually $M\gg m$ due to the budget constraint when collecting $\D$.

A representative work in MOS prediction for synthetic speech is MOSNet \cite{mosnet}, which is depicted in the left most subfigure in Figure~\ref{fig:all_models}. MOSNet aims at finding a model $f$ that predicts the subjective rating of a given speech sample. The MOSNet training involves minimizing a mean loss (using a criterion like MSE) w.r.t. the mean score of each sample:
\begin{align}
	\mathcal{L}_{\text{mean}}=\text{MSE}(f(x_i), \bar{s_i}).	 \label{eq:mean_loss}
\end{align}
During inference, given an input speech $x$, the trained model is directly used to make the prediction.
\begin{align}
	\hat{s}=f(x). \label{eq:mn_inf}
\end{align}

\subsection{MBNet}
\label{ssec:mbnet}

MBNet consists of two subnets: a MeanNet $f$ and a BiasNet $g$, as shown in the second left most subfigure in Figure~\ref{fig:all_models}. Similar to MOSnet, the MeanNet aims to predict the mean score, while the BiasNet takes not only the speech sample $x_i$ but also the listener ID $l_i^j$ as input to predict the bias score. By combining the mean score and the bias score, we obtain the LD score and the corresponding bias loss for MBNet as follows:
\begin{align}
	\mathcal{L}_{\text{bias}}=\text{MSE}(f(x_i)+g(x_i, l_i^j), s_i^j). \label{eq:bias_loss}
\end{align}
The final loss of MBNet is defined as a weighted sum of the mean loss and the bias loss:
\begin{align}
	\mathcal{L}_{\text{MBNet}}=\alpha\mathcal{L}_{\text{mean}}+\lambda\mathcal{L}_{\text{bias}}. \label{eq:mbnet_loss}
\end{align}
During inference, since we aim to assess the average subjective rating of the query sample, the BiasNet is discarded and only the MeanNet is used to make the prediction, similar to Equation~\ref{eq:mn_inf}. We refer to this inference mode as the \textit{MeanNet} inference.

\subsection{Inefficient design of MBNet}
\label{sssec:inefficiency}

Although MBNet was shown to be effective \cite{mbnet, s3r-mos}, we argue that the MBNet design is somewhat inefficient. In the original MBNet, the MeanNet and the BiasNet both take the speech as input. which raises several problems. First, there are certain criterion and standards that are invariant to listeners when rating a speech sample. In other words, the two subnets should share some common behaviors. Second, from the inference point of view, if only the MeanNet is used, then the BiasNet and the subsequent bias loss act only like a regularization. However, since the BiasNet also takes the speech as input, it is necessary but inefficient to make the BiasNet big. In a nutshell, it is worthwhile to redesign the model in order to learn \textit{listenr-dependent} and \textit{listenr-independent} feature representations.

%

\section{Listener-dependent network (LDNet)}

We first present a more general formulation of LD modeling. Consider the model structure depicted in the second right most subfigure in Figure~\ref{fig:all_models}. From the input/output perspective, we only define one single model $f$ to produce the LD score given the speech and the listener ID as input. We name our formulation \textit{LDNet} in contrast to \textit{MBNet} since we do not explicitly define two submodules to predict the mean and bias scores. During training, the model is trained only to minimize a LD loss as follows:
\begin{align}
	\mathcal{L}_{\text{LDNet}}=\mathcal{L}_{\text{LD}}=\text{MSE}(f(x_i, l_i^j), s_i^j). \label{eq:ld_loss}
\end{align}
Note that LDNet can be viewed as a generalization of MBNet, as the MBNet outputs the LD score by adding the outputs of the MeanNet and the BiasNet.

\subsection{All listeners inference}
\label{ssec:all-lis}

We then show how to perform inference with only the LDNet. Inspired by \cite{ld-ser}, we propose the \textit{all listeners} inference, which simulates the decisions of each training listener and average over them:
\begin{align}
	\hat{s}=\frac{1}{M}\sum_{j=1}^{M}f(x, l^j). \label{eq:all_lis}
\end{align}
An obvious advantage of the all listeners inference is its flexibility. Unlike the MeanNet inference defined in Equation~\ref{eq:mn_inf}, which requires an explicit network to produce the mean score, the all listeners inference mode only requires the model to be able to produce LD scores w.r.t. training listeners. That is to say, all listeners inference also applies to MBNet.

\subsection{Listener independent and dependent model decomposition}
\label{ssec:decomposition}

As we argued in Section~\ref{sssec:inefficiency}, we propose to decompose the model $f$ into an \textit{encoder} that learn listener-independent (LI) features and a \textit{decoder} that fuses the listener information to generate LD features, as depicted in Figure~\ref{fig:all_models}. Formally, we can write
\begin{align}
	f=\text{Decoder}(\text{Encoder}(x), l).
\end{align}
The division of the encoder and the decoder is simply where the listener ID is injected. That is to say, the BiasNet in MBNet can in fact be factorized in the same fashion. However, in MBNet, the encoder was a single convolutional 2D (conv2d) layer, while the decoder was much deeper. We argue that if the listener preference only adds a shift to the mean score (as the name ``bias'' suggests), then the decoder should be made simple and leave most of the representative power to the encoder, as done in \cite{ld-ser}. We present how we achieve this in our model design in later sections.

\subsection{Utilizing the mean score with a MeanNet}

We can utilize the mean scores with a MeanNet, as depicted in the rightmost subfigure in Figure~\ref{fig:all_models}. We refer to this model as LDNet-MN. Instead of taking the input speech as input, the MeanNet here takes the LI features extracted with the encoder to predict the mean score. The motivation is to help the encoder extract LI features since the mean score is LI by our assumption. This multitask learning (MTL) loss can be derived by rewriting Equation~\ref{eq:mean_loss}:
\begin{align}
	\mathcal{L}_{\text{MTL}}=\text{MSE}(\text{MeanNet}(\text{Encoder}(x)), \bar{s_i}),	 \label{eq:mean_mtl_loss}
\end{align}
and the loss of LDNet-MN can be written as:
\begin{align}
	\mathcal{L}_{\text{LDNet{-}MN}}=\alpha\mathcal{L}_{\text{MTL}}+\lambda\mathcal{L}_{\text{LD}}. \label{eq:ldnetmn_loss}
\end{align}
LDNet-MN and MBNet might appear to have a similar structure and objective, but a fundamental difference is that the MTL loss propagates back to not only the MeanNet but also the encoder, while the mean loss in Equation~\ref{eq:mbnet_loss} only affects the MeanNet.
In addition, similar to the principle described in Section~\ref{ssec:decomposition}, we designed the MTL head to be as simple as possible.

\subsection{Utilizing the mean score with a mean listener}

One shortcoming of the all listeners inference is that it requires to run multiple forward passes first and average the results. A work-around is using matrix representation to run only one forward pass, with the cost of extra memory consumption. Alternatively, we can extend the training set by adding a virtual "mean listener" (ML). Formally, each sample $x_i$ now has $m+1$ LD scores, $\{s_i^1,\dots ,s_i^m,\bar{s_i}\}$, and the listener ID that corresponds to the mean scores of each speech sample is the mean listener. We can then train a LDNet with the extended training set, and denote such a variant as LDNet-ML. Note that we did not assign a different weight (or use techniques like oversampling) when updating the model. During test time, in addition to the all listeners inference, LDNet-ML provides an efficient \textit{mean listener} inference mode, which is to simply use the mean listener ID to run one forward pass. 

\section{Experimental settings}

\subsection{Datasets}

\noindent{\textbf{VCC2018}} \cite{vcc2018} This dataset contains 20580 speech samples, where each sample was rated by 4 listeners. A total of 270 listeners were recruited. and each listener rated on average 226 samples. We followed an open-source MBNet implementation\footnote{\url{https://github.com/sky1456723/Pytorch-MBNet/}} and used a random split of 13580/3000/4000 for train/valid/test. Note that all listeners are seen listeners during validation and testing.

\noindent{\textbf{BVCC}} \cite{BVCC} This is a newly collected large-scale MOS dataset, containing samples from the past BCs and VCCs as well as state-of-the-art TTS systems implemented in ESPNet \cite{espnet, espnet-tts}.
There are 7106 samples, with each sample rated by 8 listeners. In total there are 304 listeners, with each listener rating 187 samples. A carefully curated rule was used to create a 4974/1066/1066 train/valid/test data split. There are 288 listeners in the training set, and there are 8 unseen listeners in the valid and test sets, with some overlap between the training listeners. For more details about how the split, please refer to \cite{generalization-mos}.


\subsection{Implementation}

\begin{figure}[t]
	\centering
	\includegraphics[width=\columnwidth]{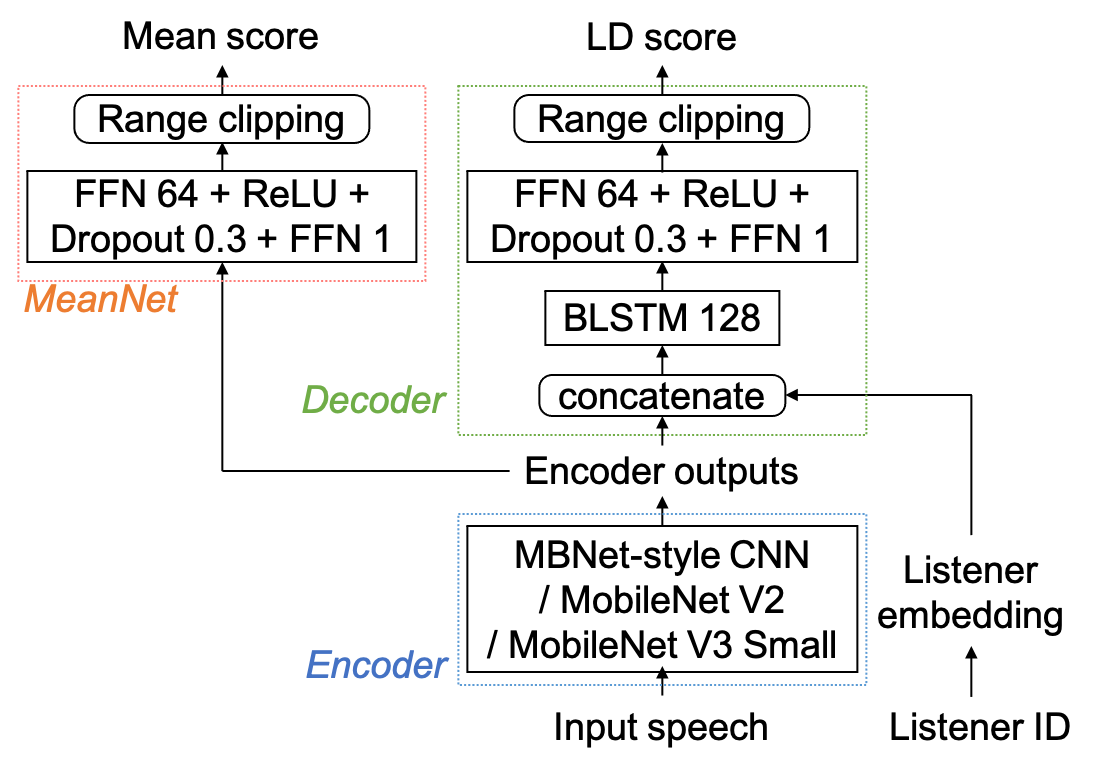}
	\centering
	\captionof{figure}{Illustration of the LDNet model architecture with a RNN-based decoder.}
	\label{fig:arch}
\end{figure}

We will open-source our codebase in the near future\footnote{\url{https://github.com/unilight/LDNet}}.

\noindent{\textbf{Baselines.}}
We used the pretrained model in the official MOSNet implementation\footnote{\url{https://github.com/lochenchou/MOSNet}}. We used an unofficial, open-source implementation of MBNet as mentioned in the previous subsection. Note that we provided our self-implemented MBNet results because the data split of the MBNet paper was not specified by the authors, so their results are not directly comparable with our LDNet results.

\noindent{\textbf{Model details.}}
The input of the models is the magnitude spectrum, which was also used in \cite{mosnet}. All models output frame-level scores, and a simple average pooling was used to generate the utterance-level scores.
For the LDNets, we tried three encoder variants. To align with MBNet, we first used the MeanNet structure in MBNet, which was composed of conv2d blocks with dropout and batch normalization layers. We then tried two efficient but powerful conv2d-based architectures, MobileNetV2 \cite{mobilenetv2} and MobileNetV3 \cite{mobilenetv3}. We used implementations provided by torchvision\footnote{\url{https://github.com/pytorch/vision/tree/main/torchvision/models}} and we refer the readers to the original papers for more details. The base structure of the decoder and the MeanNet (in LDNet-MN) is a single-layered feed-forward network (FFN) followed by projection, and we additionally experimented with a BLSTM-based RNN decoder. Figure~\ref{fig:arch} illustrates the LDNet model architecture.

\noindent{\textbf{Training details.}}
MBNet-style models were trained with the Adam optimizer with learning rate (LR) $0.001$, and MobileNet V2 and V3 models were trained using RMSprop. With V2 the LR was decayed by 0.9 every 5k steps, and with V3 the LR was decayed by 0.97 every 1k steps. The $\alpha$ and $\lambda$ in Equations~\ref{eq:mbnet_loss} and~\ref{eq:mean_mtl_loss} were set to 1 and 4, respectively.
We also used techniques from recent papers which we found helpful in our experiments. The clipped MSE \cite{mbnet} was used to prevent the model from overfitting. Repetitive padding \cite{mbnet} was found to be better than zero padding with a masked loss. Range clipping \cite{s3r-mos} was an effective inductive bias to force the range of the network output.

%

\begin{table*}[t]
	\centering
	\caption{Results on the VCC2018 test set and the BVCC test set. "MN", "All" and "ML" stand for mean net, all listeners and mean listener inference, respectively. For MSE, the smaller the better; for LCC and SRCC, the larger the better. }
	\scriptsize
	
	\centering
	\hspace*{-0.3cm}
	\begin{tabular}{ l l c c | c c c | c c c | c c c | c c c }
		\toprule
		\multirow{3}{*}[-1pt]{Model} & \multirow{3}{*}[-1pt]{\makecell{Config\\(Enc./Dec./MeanNet)}} & \multirow{3}{*}[-1pt]{\makecell{Model\\size}} & \multirow{3}{*}[-1pt]{Mode} & \multicolumn{6}{c|}{VCC2018} & \multicolumn{6}{c}{BVCC} \\
		& & & & \multicolumn{3}{c}{Utterance level} & \multicolumn{3}{c|}{System level} & \multicolumn{3}{c}{Utterance level} & \multicolumn{3}{c}{System level} \\
		& & & & MSE & LCC & SRCC & MSE & LCC & SRCC & MSE & LCC & SRCC & MSE & LCC & SRCC \\
		\midrule
		MOSNet & Numbers from \cite{mosnet} & - & MN & 0.538 & 0.642 & 0.589 & 0.084 & 0.957 & 0.888 & - & - & - & - & - & - \\
		MOSNet & Numbers from \cite{mbnet} & - & MN & 0.465 & 0.638 & 0.611 & 0.047 & 0.964 & 0.922 & - & - & - & - & - & - \\
		MOSNet & Model from \cite{mosnet} & - & MN & - & - & - & - & - & - & 0.816 & 0.294 & 0.263 & 0.563 & 0.261 & 0.266 \\
		MBNet$\dagger$ & Numbers from \cite{mbnet} & - & MN & \textbf{0.426} & \textbf{0.680} & \textbf{0.647} & 0.029 & 0.977 & 0.949 & - & - & - & - & - & - \\
		
		\midrule
		
		\multirow{2}{*}[0pt]{(a) MBNet} & \multirow{2}{*}[0pt]{Self implementation} & \multirow{2}{*}[0pt]{1.38M} &  MN & 0.955 & 0.658 & 0.630 & 0.549 & 0.978 & 0.957 & 0.669 & 0.757 & 0.765 & 0.522 & 0.854 & 0.860 \\
		& & & All & 0.615 & 0.656 & 0.627 & 0.154 & 0.980 & 0.966 & 0.492 & 0.758 & 0.765 & 0.271 & 0.856 & 0.860 \\
		(b) LDNet & MBNet-style/RNN/- & 1.18M & All & 0.465 & 0.650 & 0.617 & 0.040 & 0.973 & 0.955 & 0.397 & 0.740 & 0.734 & 0.189 & 0.856 & 0.855 \\
		(c) LDNet & MobileV2/RNN/- & 1.73M & All & 0.461 & 0.646 & 0.603 & 0.037 & 0.984 & 0.958 & 0.328 & 0.793 & 0.791 & 0.179 & 0.878 & 0.876 \\
		(d) LDNet & MobileV3/RNN/- & 1.48M & All & 0.432 & 0.676 & 0.641 & 0.020 & \textbf{0.989} & 0.976 & 0.324 & 0.794 & 0.790 & 0.174 & 0.876 & 0.871 \\
		(e) LDNet & MobileV3/FFN/- & 0.96M & All & 0.457 & 0.661 & 0.621 & \textbf{0.013} & 0.988 & 0.976 & 0.333 & 0.788 & 0.784 & 0.173 & 0.876 & 0.870 \\
		(f) LDNet-MN & MobileV3/RNN/FFN & 1.49M & All & 0.437 & 0.671 & 0.635 & 0.023 & 0.987 & 0.971 & 0.324 & 0.794 & 0.791 & 0.187 & 0.869 & 0.868 \\
		\multirow{2}{*}[0pt]{(g) LDNet-ML} & \multirow{2}{*}[0pt]{MobileV3/FNN/-} & \multirow{2}{*}[0pt]{0.96M} & All & 0.463 & 0.653 & 0.617 & 0.024 & 0.983 & 0.975 & \textbf{0.316} & \textbf{0.795} & \textbf{0.794} & \textbf{0.157} & 0.881 & 0.881 \\
		& & & ML & 0.479 & 0.648 & 0.613 & 0.021 & 0.983 & \textbf{0.979} & 0.333 & \textbf{0.795} & \textbf{0.794} & 0.169 & \textbf{0.885} & \textbf{0.886} \\
		\bottomrule
		\multicolumn{16}{l}{$\dagger$: The results on this row are not directly comparable with the rows below since it is unclear what data split the authors used. We suggest readers to compare results from (a) to (g).}
	\end{tabular}
	\label{tab:main-results}
	\vspace*{-0.3cm}
\end{table*}

\section{Experimental results}

For all self implemented models, we used 3 different random seeds and report the averaged metrics including MSE, linear correlation coeffieient (LCC) and Spearsmans rank correlation coeffieient (SRCC) in both utterance level and system level. The MBNet and LDNets were trained for 50k and 100k steps, respectively, and following \cite{s3r-mos}, model selection was based on system-level SRCC.


Our main experimental results are shown in Table~\ref{tab:main-results}.  We summarize our observations into the following points.

\subsection{Advantage of all listeners inference}

As mentioned in Section~\ref{ssec:all-lis}, all listeners inference can be applied to any model that produces LD scores w.r.t. training listeners, including MBNet. In row (a), compared to mean net inference, all listeners inference greatly reduces both utterance and system level MSE, and provided a slight system-level improvement. This shows that all listeners inference can reduce the variance of the prediction and better capture the relationship between different systems.

\subsection{Impact of encoder design}

We then investigate the impact of the encoder design by comparing rows (b), (c) and (d) which used the MBNet-style encoder, MobileNetV2 and MobileNetV3, respectively. On VCC2018, we observes an stable system-level improvement as the encoder advances. On BVCC, although the MobileNetV3 encoder gave the lowest system-level MSE, its LCC and SRCC were slightly lower than those of the MobileNetV2 encoder. Considering that MobileNetV3 used 0.25M less model parameters, and empirically its training time was $33\%$ faster than that of MobileNetV2, we fixed the encoder to MobileNetV3 in succeeding experiments.

\subsection{Impact of decoder design} 

The influence of a simpler decoder design was inspected by removing the RNN layer in model (d) to form a simple FFN decoder. The resulting model (e) had a comparable system level performance, a $64\%$ reduction in model size and an empirical $25\%$ faster training time. This result is consistent with our argument in Section~\ref{ssec:decomposition} that the decoder can be made as simple as possible as long as we have a encoder which is strong enough.

\subsection{Effectiveness of LDNet-MN}

LDNet-MN shared a similar structure with MBNet and was expected to bring improvement by utilizing the mean score. However, by comparing rows (d) and (f), we observed no improvements but only degradation in both utterance-level and system-level. We also tried an RNN-based MeanNet but still observed no improvements.

\subsection{Effectiveness of LDNet-ML}

We finally examined LDNet-ML, which utilized the mean score in the listener embedding space. By comparing rows (e) and (g), when both using all listeners inference, a slight degradation was observed on VCC2018, while a substantial improvement was observed on BVCC. Interestingly, when switched to mean listener inference, further improvements on the system level SRCC can be obtained on both datasets. There results suggest that LDNet-ML is a better way to utilize the mean score than LDNet-MN. Also, since VCC2018 has 4 ratings per sample and BVCC has 8, the mean score in BVCC is considered more reliable, resulting in more significant improvements brought by LDNet-ML.

\section{Conclusions and future works}

In this work, we integrated recent advances in LD modeling for MOS prediction. The resulting model, LDNet, was equipped with an advanced model structure and several inference options for efficient prediction.
Evaluation results justified the design of the proposed components and showed that our system outperformed the MBNet baselines.
Results also showed that LDNet-ML is the best way to utilize the mean scores, and its advantage is even more prevailing when we have more ratings per sample.
In the future, as the proposed techniques are flexible, we plan to combine them with existing effective methods, such as time-domain modeling \cite{time-domain-nisqa, metricnet} and S3R learning \cite{s3r-mos}.

\vspace*{-1.5mm}
\section{Acknowledgements}

The authors would like to thank the organizers of the Blizzard Challenges and Voice Conversion Challenges for providing the data. This work was partly supported by JSPS KAKENHI Grant Number 21J20920 and JST CREST Grant Number JPMJCR19A3, Japan. This work was also supported  by  JST  CREST  grants  JPMJCR18A6 and JPMJCR20D3, and by MEXT KAKENHI grants 21K11951 and 21K19808.

\bibliographystyle{IEEEbib}
\bibliography{ref}

\end{document}